\documentclass[letterpaper,11pt]{article}
\usepackage{amsmath}
\usepackage{amssymb}
\usepackage[dvips]{graphicx}
\begin{document}
\title{Analytical potential-density pairs from complex-shifted Kuzmin-Toomre discs}
\author{D. Vogt\thanks{e-mail: dvogt@ime.unicamp.br} 
\and
P. S. Letelier\thanks{e-mail: letelier@ime.unicamp.br}\\
Departamento de Matem\'{a}tica Aplicada-IMECC, Universidade \\ 
Estadual de Campinas 13083-970 Campinas, S.\ P., Brazil}
\maketitle
\begin{abstract}
The complex-shift method is applied to the Kuzmin-Toomre family of discs 
to generate a family of non-axisymmetric flat distributions of matter. 
These are then superposed to construct non-axisymmetric flat rings. We 
also consider triaxial potential-density pairs obtained from
these non-axisymmetric flat systems by means of suitable transformations.
The use of the imaginary part of complex-shifted potential-density 
pairs is also discussed. 

\end{abstract}
\section{Introduction}

One of the difficulties encountered in modelling self-gravitating systems 
is set by the potential theory. In general, to calculate the gravitational potential 
associated with a matter distribution of arbitrary shape, one has to use 
numerical methods. On the other hand, it is convenient to have a set of
analytical potential-density pairs that provide a description
of gravitating systems, in particular, for those that deviate from spherical symmetry. 
A survey of such models can be found in \cite{bt08}, chapter 2. 

A simple way to construct axisymmetric potential-density pairs
that represent discs is the image method that is usually used to solve 
problems in electrostatics.
This method was employed in the context of the Newtonian gravity first by Kuzmin \cite{k56}, who
derived a very simple potential-denstity pair of a disc (Section \ref{sec_disc}).
Several years later, Evans \& de Zeeuw \cite{ez92} showed that
any axisymmetric disc can be constructed by superposing simple Kuzmin discs with different
weights. The image method has also been adapted to and used in General Relativity to generate several
exact solutions of Einstein's equations that represent discs (see e.g.\ \cite{blk92,blp93,lz99,let99,gl00,vl03,gl04,vl05}).

Another technique to construct new solutions in the Newtonian gravity was introduced by Appell \cite{ap87,ww50} who 
considered the complexification of the potential of a point mass. This method, known as \emph{complex-shift method},
was also used in General Relativity \cite{lo87,gp89,lo98}, but see some misinterpretations in the Appendix of \cite{dlo05}. 
More recently, the complex-shift method has been applied to spherical and axisymmetric Newtonian parent systems to generate 
new analytical axisymmetric and triaxial potential-density pairs \cite{cg07,cm08}. 

In this work, we use the complex-shift method to derive potential-dentity pairs for 
non-axisymmetric flat discs and rings and their ``inflated'' versions. 
In Section \ref{sec_complex}, we summarize the method of complexification. In
Section \ref{sec_disc}, we complex shift the Kuzmin-Toomre family of discs to derive 
a family of non-axisymmetric flat discs. The members of this class are then 
superposed to generate further non-axisymmetric flat distributions of matter 
that can be interpreted as complex-shifted rings. This is done in Section \ref{sec_ring}. 
In Section \ref{sec_tri_disc}, we use a general transformation to ``inflate'' the
previous non-axisymmetric flat systems to generate triaxial potential-density pairs. 
We derive a generalized pair that includes an example studied by Ciotti \& Marinacci \cite{cm08}.
In Section \ref{sec_im}, we show that the imaginary part of a complex-shifted 
system can be superposed to another potential-density pair to generate physically 
acceptable systems which are not symmetrical with respect to a coordinate axis. 
At last, in Section \ref{sec_discuss} we summarize our results.
\section{The complex-shift method} \label{sec_complex}

In this section, we briefly expose the idea of the complex-shift method. The formalism
is similar to the presented by Ciotti \& Giampieri \cite{cg07} and Ciotti \& Marinacci \cite{cm08}.
We start by considering a gravitational potential $\Phi(\mathbf{x})$ and an associated
density distribution $\rho(\mathbf{x})$, where $\mathbf{x}=(x,y,z)$ is the position
vector. We also assume that $\rho(\mathbf{x})$ satisfies the Poisson equation 
\begin{equation} \label{eq_poisson} 
\nabla^2\Phi=4\pi G \rho \mbox{.}
\end{equation}
The basic idea of the complex-shift method is to obtain new potential-density pairs
by a displacement of a given potential-density pair (the parent system) on the imaginary axis. The complexified
potential is defined as $\Phi_c=\Phi(\mathbf{x}-i\mathbf{a})$, where $i$ is the 
imaginary unit and $\mathbf{a}$ is a real shift vector. Since the complex shift
is a linear transformation and the Poisson equation is a linear partial differential equation (PDE) acting on 
the $\mathbf{x}$ vector, it follows that
\begin{equation} 
\nabla^2 \Phi_c=4\pi G \rho_c \mbox{,}
\end{equation}
where $\rho_c=\rho(\mathbf{x}-i\mathbf{a})$ is the shifted mass density. By separating 
the real and imaginary parts of $\Phi_c$ and $\rho_c$, one obtains two real 
potential-density pairs. 

Unfortunately, there is no guarantee that the resulting pairs describe physically acceptable 
(i.e.\ everywhere positive) gravitating systems. In fact, Ciotti \& Giampieri \cite{cg07} have shown that the imaginary part of
the complex-shifted density always changes sign because the total mass of the 
complexified potential-density pair coincides with the total mass of the seed 
density distribution, and thus $\int \Im (\rho_c) \mathrm{d}^3\mathbf{x}=0$. 
However, the real part can be positive everywhere. This usually happens for 
a finite domain of the shift vector $\mathbf{a}$.
\section{Thin shifted Kuzmin-Toomre discs} \label{sec_disc}

In this Section, we apply the complex shift to the Kuzmin-Toomre family of 
discs \cite{k56,t63} to generate flat distributions of matter without axial symmetry.
We begin with the first member of this family, which was the simple model derived by Kuzmin. 
The potential in the usual cylindrical coordinates $(R,z,\varphi)$  is given by
\begin{equation} \label{eq_phi_kuz}
\Phi=-\frac{GM}{\sqrt{R^2+(b+|z|)^2}} \mbox{,}
\end{equation}
where $b$ is a non-negative parameter. Using Poisson equation (\ref{eq_poisson}),
the discontinuous normal derivative on $z=0$ introduces a surface density of mass
\begin{equation} \label{eq_sigma}
\sigma(R)=\frac{1}{2\pi G}\left. \frac{\partial \Phi}{\partial z}\right|_{z=0} \mbox{,}
\end{equation}
which for the potential (\ref{eq_phi_kuz}) results
\begin{equation}  \label{eq_sigma_kuz}
\sigma(R)=\frac{bM}{2\pi \left( R^2+b^2 \right)^{3/2}} \mbox{.}
\end{equation}
The result of a complex shift of an axisymmetric potential depends not only on 
the length but also on the direction
of the shift vector $\mathbf{a}$. We are interested in the equatorial shift
of the potential (\ref{eq_phi_kuz}), where without loss of generality we make a 
shift along the $x$-axis, $\mathbf{a}=(a,0,0)$. The complexified potential
(\ref{eq_phi_kuz}) reads
\begin{equation} \label{eq_kuz_c}
\Phi_c=-\frac{GM}{\sqrt{(x-ia)^2+y^2+(b+|z|)^2}} \mbox{.}
\end{equation}
To evaluate the square root, we denote $d^2=x^2+y^2-a^2+(b+|z|)^2-2iax=u\mathrm{e}^{i\theta}$. Thus,
\begin{align}
u =|d|^2 &=\sqrt{\left[ x^2+y^2-a^2+(b+|z|)^2 \right]^2+4a^2x^2} \mbox{,} \label{eq_u}\\
\cos \theta &=\frac{x^2+y^2-a^2+(b+|z|)^2}{u}, \qquad \sin \theta =-\frac{2ax}{u} \mbox{.} \label{eq_theta}
\end{align}
For simplicity, we restrict to the case $\cos \theta >0$ everywhere, so we must have $a<b$. This restriction will
be assumed henceforth. To make the square root a single valued function, we choose
$d=\sqrt{u}\mathrm{e}^{i\theta/2}$, with 
\begin{equation} \label{eq_dupl}
\cos \frac{\theta}{2}=\sqrt{\frac{1+\cos \theta}{2}}, \qquad \sin \frac{\theta}{2}=
\frac{\sin \theta}{2\cos (\theta/2)} \mbox{.}
\end{equation}
With the help of equations (\ref{eq_u})--(\ref{eq_dupl}), the real and imaginary 
parts of the potential (\ref{eq_kuz_c}) can be expressed as 
\begin{gather}
\Re(\Phi_c) =-\frac{GM}{\sqrt{2}}\sqrt{\frac{\chi+\sqrt{\chi^2+4a^2x^2}}{\chi^2+4a^2x^2}} \mbox{,} \label{eq_re_pk} \\
\Im(\Phi_c) =-\frac{\sqrt{2}GMax}{\sqrt{\chi^2+4a^2x^2}\sqrt{\chi+\sqrt{\chi^2+4a^2x^2}}} \mbox{,} \label{eq_im_pk}
\end{gather}
where $\chi=x^2+y^2-a^2+(b+|z|)^2$. The respective surface densities can be calculated
using relation (\ref{eq_sigma}) 
\begin{gather}
\Re(\sigma_c) =\frac{Mb\sqrt{\xi+\sqrt{\xi^2+4a^2x^2}}}{2 \sqrt{2} \pi 
\left(\xi^2+4a^2x^2 \right)^{3/2}} \left( 2\xi-\sqrt{\xi^2+4a^2x^2} \right) \mbox{,}
\label{eq_re_sk} \\
\Im(\sigma_c) =\frac{Mabx\left( 2\xi+\sqrt{\xi^2+4a^2x^2} \right)}{\sqrt{2} \pi 
\left(\xi^2+4a^2x^2 \right)^{3/2} \sqrt{\xi+\sqrt{\xi^2+4a^2x^2}}} \label{eq_im_sk} \mbox{,}
\end{gather}
where we defined $\xi=x^2+y^2+b^2-a^2$.
Equations (\ref{eq_re_sk}) and (\ref{eq_im_sk}) may also be obtained from the real and imaginary parts
of the complexified surface density (\ref{eq_sigma_kuz}). 

\begin{figure}
\centering
\includegraphics[scale=0.75]{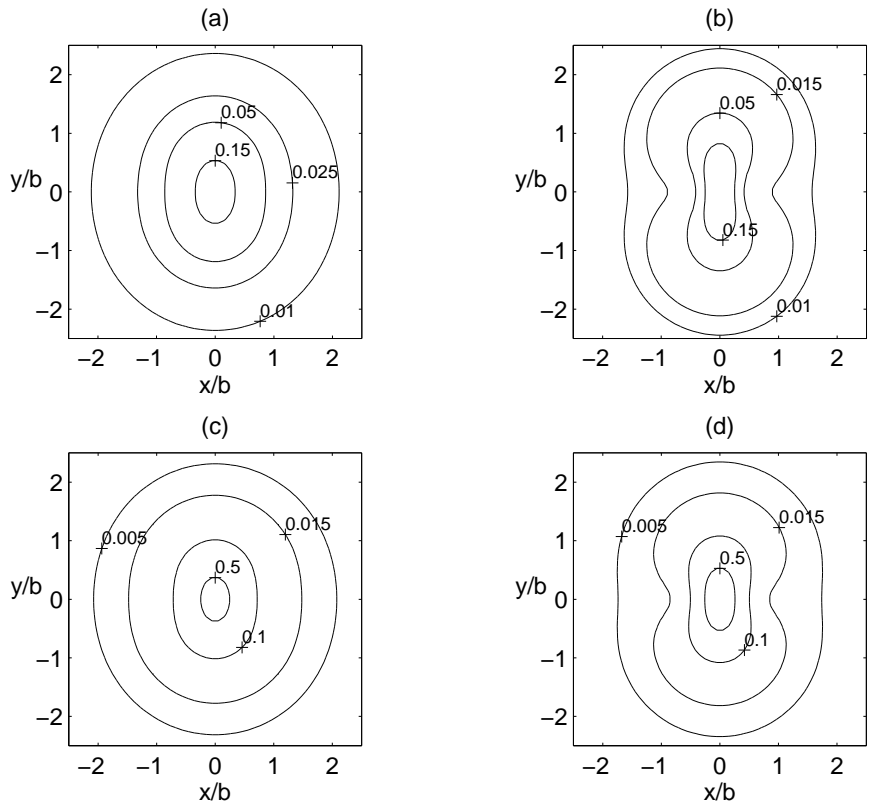}
\caption{(a) and (b) Isodensity curves of the surface density (\ref{eq_re_sk}) as functions of $x/b$ and $y/b$ 
for (a) $a/b=0.5$ and (b) $a/b=0.8$. (c) and (d) Isodensity curves of the surface density (\ref{eq_sigma1}) 
as function of $x/b$ and $y/b$ for (b) $a/b=0.4$ and (b) $a/b=0.55$.} \label{fig_disc}
\end{figure}

As was stated in Section \ref{sec_complex}, the imaginary part of the shifted Kuzmin-Toomre 
disc changes sign and becomes negative for $x<0$, so we focus on the real part. The expansion 
of the density (\ref{eq_re_sk}) about the origin yields 
\begin{equation}
\Re(\sigma_c) \approx \frac{Mb}{2\pi (b^2-a^2)^{3/2}} -\frac{3Mb(b^2+4a^2)x^2}{4\pi (b^2-a^2)^{7/2}}-
\frac{3Mby^2}{4\pi (b^2-a^2)^{5/2}} \mbox{.}  
\end{equation} 
The isodensity curves are ellipses with major axis on the $y$-axis, minor-to-major squared axis ratio 
$(b^2-a^2)/(b^2+4a^2)$ and eccentricity $e=\sqrt{5a^2/(b^2+4a^2)}$. Some curves of constant density 
in units of $M/b^2$ are displayed in Figs.\ \ref{fig_disc}(a) and (b) as functions of $x/b$ and $y/b$ 
for (a) $a/b=0.5$ and (b) $a/b=0.8$. We have a monotone decreasing surface density in the 
interval $0< a/b \lesssim 0.59$, whereas the density is always non-negative for $0< a/b \lesssim 0.86$.
Expanding the potential (\ref{eq_re_pk}) about the origin in the plane $z=0$ results 
\begin{equation} 
\Re(\Phi_c) \approx -\frac{GM}{\sqrt{b^2-a^2}} +\frac{GM(b^2+2a^2)x^2}{2(b^2-a^2)^{5/2}}
+\frac{GMy^2}{2(b^2-a^2)^{3/2}} \mbox{,} 
\end{equation} 
which is the potential of an anisotropic harmonic oscillator with the ratio of frequencies 
\begin{equation} 
\frac{\omega_x}{\omega_y}=\sqrt{\frac{b^2+2a^2}{b^2-a^2}} \mbox{.}
\end{equation} 
Therefore, we expect that test particles near the centre move in box orbits \cite{bt08}. 
The contours in Fig.\ \ref{fig_disc}(b) suggest that the simple potential-density pair 
(\ref{eq_re_pk}) and (\ref{eq_re_sk}) may be used to represent thin galactic discs with a central bar.  

The potential-density pair obtained form the real part of the shifted Kuzmin-Toomre disc is 
in fact the first member of the family of non-axisymmetric thin disc-like structures. The other members 
can be calculated by using the recurrence relations for the Kuzmin-Toomre family of discs 
\cite{nm76,blp93} 
\begin{gather}
\Phi_{c(n+1)} =\Phi_{c(n)}-\frac{b}{2n+1} \frac{\partial}{\partial b} \Phi_{c(n)} \mbox{,} \\
\sigma_{c(n+1)} =\sigma_{c(n)}-\frac{b}{2n+1} \frac{\partial}{\partial b} \sigma_{c(n)}
\text{,} \qquad n=0,1,\ldots \mbox{,}
\end{gather}
where $\Phi_{c(0)}$ and $\sigma_{c(0)}$ are  given by equations 
(\ref{eq_re_pk}) and (\ref{eq_re_sk}), respectively. For instance, the potential and density of 
the next member read 
\begin{gather}
\Phi_{c(1)} =-\frac{GM\sqrt{\chi+\sqrt{\chi^2+4a^2x^2}}}{\sqrt{2}\left( \chi^2+4a^2x^2 \right)^{3/2}} \left[ 
\chi^2+4a^2x^2  \right. \notag \\
\left. +b(b+|z|) \left(2\chi-\sqrt{\chi^2+4a^2x^2} \right) \right]  \mbox{,}  \label{eq_phi1} \\
\sigma_{c(1)} =\frac{3Mb^3\sqrt{\xi+\sqrt{\xi^2+4a^2x^2}}}{2\sqrt{2} \pi (\xi^2+4a^2x^2)^{5/2}} \left( 
3\xi^2-2\xi  \sqrt{\xi^2+4a^2x^2} -4a^2x^2 \right) \mbox{,} \label{eq_sigma1}
\end{gather} 
where $\chi=x^2+y^2-a^2+(b+|z|)^2$ and $\xi=x^2+y^2+b^2-a^2$. Some contours of the 
surface density (\ref{eq_sigma1}) are shown in Figs.\ \ref{fig_disc}(c) and (d) as functions of $x/b$ and $y/b$ 
for shift parameters (a) $a/b=0.4$ and (b) $a/b=0.55$. Now a monotone decreasing density is found in the 
range $0< a/b \lesssim 0.44$, and a non-negative density in the interval $0< a/b \lesssim 0.58$. 
\section{Non-axisymmetric thin rings} \label{sec_ring}

We now use the family of non-axisymmetric thin discs discussed in the previous section to 
construct structures that can be considered as non-axisymmetric flat rings. The idea is to use 
superpositions of different members of the Kuzmin-Toomre family of discs to generate rings 
(zero mass density at the origin), as was done recently by Vogt \& Letelier \cite{vl09}. Let us briefly recall the 
main point of that work. The general expression for the surface density of the 
\emph{n}th-order Kuzmin-Toomre disc is \cite{nm76}
\begin{equation} \label{eq_sigma_gen}
\sigma_{(n)}=\frac{(2n+1)b^{2n+1}M}{2\pi (R^2+b^2)^{n+3/2}} \text{,} \qquad n=0,1,\ldots \mbox{,}
\end{equation}
For convenience, we take discs with mass 
\begin{equation}
M=\frac{2\pi b^2\sigma_l}{2n+1} \mbox{,}
\end{equation}
where $\sigma_l$ is a constant with dimensions of surface density. Equation (\ref{eq_sigma_gen}) 
is then rewritten as 
\begin{equation} \label{eq_sigma_ad}
\sigma_{(n)}=\frac{\sigma_lb^{2n+3}}{(R^2+b^2)^{n+3/2}}=
\frac{\sigma_l}{\left(1+R^2/b^2 \right)^{n+3/2}}\mbox{.}
\end{equation}
Now we consider the following superposition:
\begin{multline} \label{eq_sigma_sum}
\sigma^{(m,n)}=\sum_{k=0}^m C^m_k(-1)^{m-k}\sigma_{(n+m-k)} \\
=\frac{\sigma_l}{\left(1+R^2/b^2 \right)^{n+m+3/2}}\sum_{k=0}^m C^m_k(-1)^{m-k}
\left(1+\frac{R^2}{b^2} \right)^k \mbox{,}
\end{multline}
where $C^m_k=m!/[(m-k)!k!]$. Using
\begin{equation}
\sum_{k=0}^m C^m_k(-1)^{m-k}\left(1+\frac{R^2}{b^2} \right)^k=\left( \frac{R}{b} \right)^{2m} \mbox{,}
\end{equation}
equation (\ref{eq_sigma_sum}) takes the form
\begin{equation} \label{eq_sigma_r1}
\sigma^{(m,n)}=\frac{\sigma_l\left( R/b \right)^{2m}}{\left(1+R^2/b^2 \right)^{n+m+3/2}} \mbox{.}
\end{equation}
We have that (\ref{eq_sigma_r1}) with $m=1,2,\ldots$ and $n=0,1,\ldots$ defines a family
of flat rings (characterized by zero density on $R=0$). 

It would seem that a complexification of the density (\ref{eq_sigma_r1}) would give 
rise to other ring structures. However, consider a shift $\sigma^{(m,n)}(x-ia)$ and evaluate the 
resulting complex density at the origin. We get 
\begin{equation}
\sigma_{c}^{(m,n)}= \frac{\sigma_l(-1)^m\left( a/b \right)^{2m}}{\left(1-a^2/b^2 \right)^{n+m+3/2}} \mbox{,}
\end{equation}   
which is non-zero and even negative when $m$ is odd. This indicates that in order to generate non-axisymmetric 
ring structures we should consider the superposition of complex-shifted Kuzmin-Toomre discs.  We go back and 
apply a shift to equation (\ref{eq_sigma_ad}), evaluate it at $(x=0,y=0)$, 
\begin{equation}
\sigma_{c(n)}=\frac{\sigma_l}{\left(1-a^2/b^2 \right)^{n+3/2}}\mbox{,}
\end{equation}
and write the superposition
\begin{equation}  \label{eq_sum_2}
\sum_{k=0}^m \frac{\alpha_k\sigma_l}{\left( 1-a^2/b^2 \right)^{m-k+n+3/2}}=
\frac{\sigma_l}{\left( 1-a^2/b^2 \right)^{m+n+3/2}} \sum_{k=0}^m \alpha_k \left( 1-\frac{a^2}{b^2} \right)^k \mbox{.}
\end{equation}
Now we impose the condition that the sum (\ref{eq_sum_2}) be zero. Noting that 
\begin{equation} \label{eq_sum_3}
\sum_{k=0}^m C^m_k \left[ -\left( 1- \frac{a^2}{b^2} \right) \right]^{m-k} \left( 1- \frac{a^2}{b^2} \right)^k=
\left( -1+\frac{a^2}{b^2}+1-\frac{a^2}{b^2} \right)^m=0 \mbox{,}
\end{equation}
and comparing with (\ref{eq_sum_2}), we get 
\begin{equation} 
\alpha_k=C^m_k(-1)^{m-k}\left( 1- \frac{a^2}{b^2} \right)^{m-k} \mbox{.}
\end{equation} 
If we denote the surface density of  the \emph{n}th-order shifted Kuzmin-Toomre disc
by $\sigma_{c(n)}$, then the superposition 
\begin{equation} \label{eq_sum_shift}
\sigma_c^{(m,n)}=\sum_{k=0}^mC^m_k(-1)^{m-k}\left( 1- \frac{a^2}{b^2} \right)^{m-k} 
\sigma_{c(n+m-k)}  \mbox{,}
\end{equation}
will be the equivalent to complex-shifted ``rings'' with zero surface density at the origin, 
and if $a=0$ the superposition reduces to the sum (\ref{eq_sigma_sum}). 
The corresponding potential is calculated by a sum similar to (\ref{eq_sum_shift}).
Note that the complex-shift of the family of rings given by (\ref{eq_sigma_r1}) and 
the superposition of complex-shifted Kuzmin-Toomre discs given by (\ref{eq_sum_shift}) 
result in different structures, because the coefficients used in the sums are different.     

\begin{figure}
\centering
\includegraphics[scale=0.75]{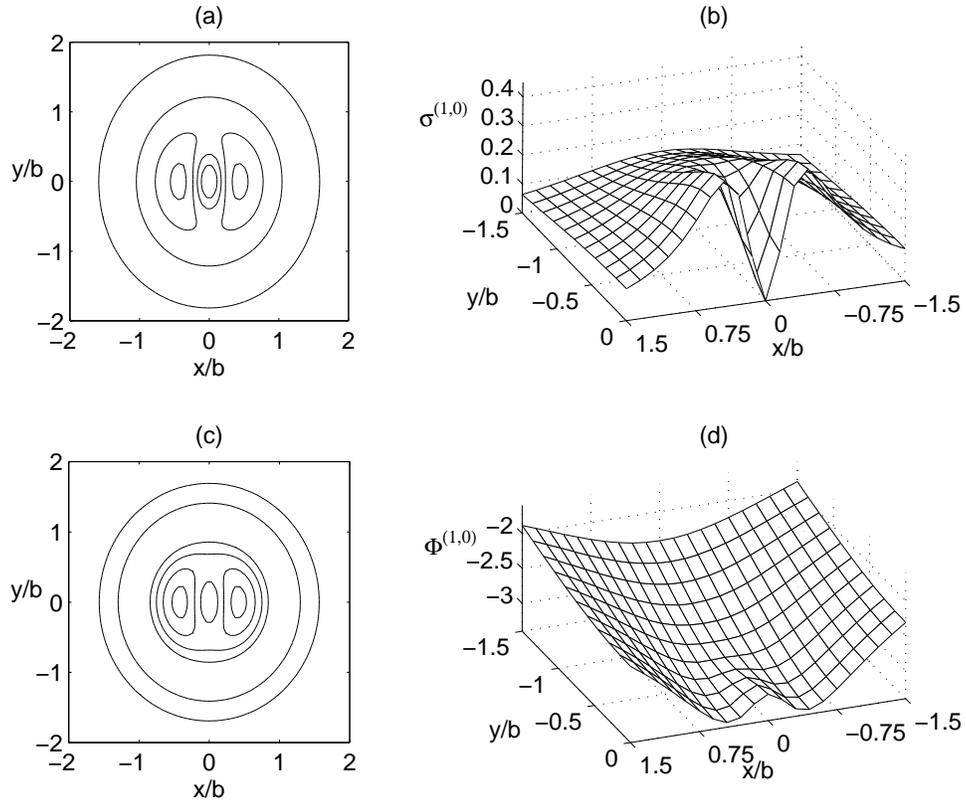}
\caption{(a) and (b) Isodensity curves and surface plot of the
 surface density $\sigma_c^{(1,0)}/\sigma_l$ equation (\ref{eq_sigma10}) with parameter $a/b=0.5$. 
(c) and (d) Isopotential curves and surface plot of the potential $\Phi_c^{(1,0)}/(\sigma_lGb)$ 
(\ref{eq_phi10}) with parameter $a/b=0.5$.} \label{fig_ring}
\end{figure}

As an example, we analyse the member with $m=1$, $n=0$
\begin{equation}
\sigma_c^{(1,0)}=\sigma_{c(0)} -\left( 1-\frac{a^2}{b^2} \right) \sigma_{c(1)} \mbox{.}
\end{equation}
Using equations (\ref{eq_re_pk}), (\ref{eq_re_sk}) and (\ref{eq_phi1})--(\ref{eq_sigma1}), the 
explicit expressions for the surface density and potential can be cast as 
\begin{gather} 
\sigma_c^{(1,0)} =\frac{\sigma_lb^3\sqrt{\xi+\sqrt{\xi^2+4a^2x^2}}}{\sqrt{2}(\xi^2+4a^2x^2)^{5/2}}
\left[ \left( \xi^2+4a^2x^2 \right) \left( 2\xi-\sqrt{\xi^2+4a^2x^2} \right) \right. \notag \\
\left. -(b^2-a^2) \left( 
3\xi^2-2\xi  \sqrt{\xi^2+4a^2x^2} -4a^2x^2 \right) \right] \mbox{,} \label{eq_sigma10}  \\
\Phi_c^{(1,0)} =-\frac{2\pi \sigma_l Gb^2\sqrt{\chi+\sqrt{\chi^2+4a^2x^2}}}
{3\sqrt{2}\left( \chi^2+4a^2x^2 \right)^{3/2}} \left[ \left( 2+\frac{a^2}{b^2} \right) 
(\chi^2+4a^2x^2) \right. \notag \\
\left. -\left( 1-\frac{a^2}{b^2} \right)b(b+|z|) \left( 2\chi - \sqrt{\chi^2+4a^2x^2} \right) 
\right] \mbox{,} \label{eq_phi10}
\end{gather}
where $\xi$ and $\chi$ were defined in Section \ref{sec_disc}. Near the centre, the approximate 
expressions for the potential-density pair are 
\begin{gather} 
\sigma_c^{(1,0)}  \approx \frac{\sigma_lb^3(b^2+9a^2)x^2}{(b^2-a^2)^{7/2}}+
\frac{\sigma_lb^3y^2}{(b^2-a^2)^{5/2}} \mbox{,}   \\
\Phi_c^{(1,0)}  \approx -\frac{2\pi \sigma_l G(b^2+a^2)}{3(b^2-a^2)^{1/2}}-
\frac{\pi \sigma_l G(b^4+7a^2b^2-2a^4)x^2}{3(b^2-a^2)^{5/2}}-
\frac{\pi \sigma_l Gy^2}{3(b^2-a^2)^{1/2}} \mbox{,}
\end{gather}
which show that the density (potential) has a minimum (maximum) at the centre.  
Figs.\ \ref{fig_ring}(a)--(d) show level curves and surface plots of the dimensionless 
density $\sigma_c^{(1,0)}/\sigma_l$ and potential $\Phi_c^{(1,0)}/(\sigma_lGb)$ with shift parameter
$a/b=0.5$. Note that we do not have exactly a homogeneous mass concentration along a ring. Two points 
of maximum occur at $(x=\pm x_s,y=0)$, where $x_s$ is the non-zero root of the equation 
$\partial \sigma_{c}^{(1,0)}/\partial x=0$ and has to be found numerically. Two saddle
points also occur at $(x=0,y=\pm \sqrt{2(b^2-a^2)/3})$. For $a/b=0.5$, we find $x_s/b \approx 0.43$ and
saddle points at $y/b=\pm \sqrt{2}/2 \approx \pm 0.71$. The potential shows two minimum points 
at $(x=\pm x_p,y=0)$, where $x_p$ is the non-zero root of the equation 
$\partial \Phi_{c}^{(1,0)}/\partial x=0$, and two saddle points at $(x=0,y=\pm (b^2-a^2)/\sqrt{a^2+2b^2})$. 
For $a/b=0.5$, we find $x_p/b \approx 0.42$ and saddle points at $y/b=\pm 1/2$.

The family of potential-density pairs derived in this Section may be useful to 
model ring galaxies, or R galaxies \cite{ts76}, objects in the form of
approximate elliptical rings; in particular a subclass of such galaxies, 
designated RE galaxies, which have crisp, elliptical rings with photographically empty interiors. 
\section{Triaxial potential-density pairs from complex-shifted discs} \label{sec_tri_disc}

The systems studied in previous sections were all flat, i.e., had infinitesimal thickness. 
One way to construct three-dimensional potential-density pairs from flat  
gravitating systems is to apply a transformation that involves the $z$-coordinate. For instance,
Miyamoto \& Nagai \cite{mn75} proposed a transformation to ``inflate'' the thin Kuzmin-Toomre
discs to obtain three-dimensional potential-density pairs for the disc part of galaxies.
Other transformation based on a class of even polynomials was used
by Gonz\'alez \& Letelier \cite{gl04} and Vogt \& Letelier \cite{vl05} to generate thick General Relativistic discs.

We shall denote by $h(z)$ a general even function of the $z$-coordinate, and start with 
the real part of the potential of the complex-shifted Kuzmin-Toomre disc, where we replace $|z|$ by $h(z)$
\begin{equation}  \label{eq_phi_c0} 
 \Re(\Phi_c) =-\frac{GM}{\sqrt{2}}\sqrt{\frac{\zeta+\sqrt{\zeta^2+4a^2x^2}}{\zeta^2+4a^2x^2}} \mbox{,} 
\end{equation}
where now we define $\zeta=x^2+y^2-a^2+[b+h(z)]^2$. The resulting mass
density follows directly from Poisson equation (\ref{eq_poisson}) in the usual Cartesian 
coordinates. We obtain
\begin{multline} 
\rho_{c(0)}=\frac{M(b+h)h_{,zz}\sqrt{\zeta+\sqrt{\zeta^2+4a^2x^2}}}{4\sqrt{2}\pi \left( \zeta^2+4a^2x^2 \right)^{3/2}}
\left( 2\zeta - \sqrt{\zeta^2+4a^2x^2} \right) \\
+\frac{M\left(1-h_{,z}^2 \right)\sqrt{\zeta+\sqrt{\zeta^2+4a^2x^2}}}{4\sqrt{2}\pi \left( \zeta^2+4a^2x^2 \right)^{5/2}}
\left[ 3(b+h)^2 \right. \\
\left. \times \left( 3\zeta^2-2\zeta  \sqrt{\zeta^2+4a^2x^2} -4a^2x^2 \right) 
-\left( \zeta^2+4a^2x^2 \right) \left( 2\zeta - \sqrt{\zeta^2+4a^2x^2} \right) \right] \mbox{.}  \label{eq_rho_c0}
\end{multline}
\begin{figure}
\centering
\includegraphics[scale=0.75]{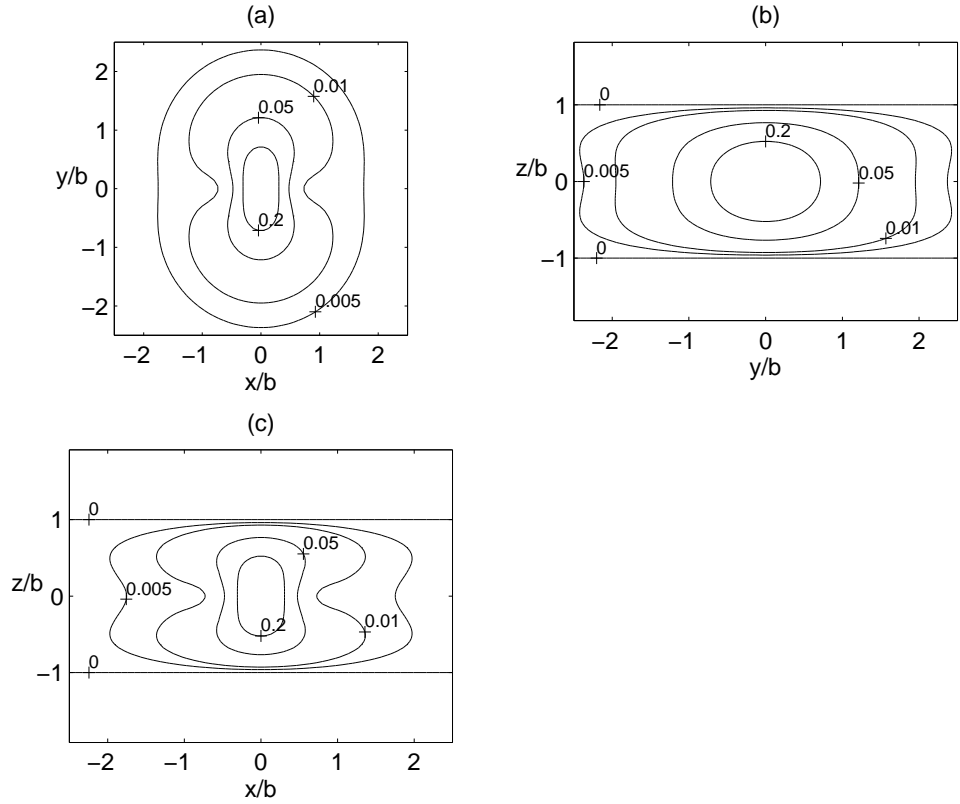}
\caption{Isodensity curves of the mass density $\rho_{c(0)}/(M/b^3)$ equation (\ref{eq_rho_c0}) 
in the three orthogonal coordinate planes for the polynomial class of functions (\ref{eq_h}). 
Parameters: $a/b=0.6$, $n=1$, $w=0$ and $s/b=1$.} \label{fig_rho_c0}
\end{figure}
Note that to satisfy the restriction 
$\cos \theta >0$ in (\ref{eq_theta}), we must have $\zeta>0$ everywhere.  
In the case of a Miyamoto \& Nagai  transformation \cite{mn75}, $h(z)=\sqrt{z^2+c^2}$ with $c>0$, and the
potential-density pair (\ref{eq_phi_c0}) and (\ref{eq_rho_c0}) is the same as studied by Ciotti \& Marinacci 
\cite{cm08}, who performed an equatorial shift of the Miyamoto-Nagai disc. We can also 
use the class of polynomials used in \cite{gl04,vl05} 
\begin{equation} \label{eq_h}
h(z)= \begin{cases}
-z+C, & z \leq -s, \\
Az^2+Bz^{2n+2}, & -s \leq z \leq s, \\
z+C, & z \geq s,
\end{cases}
\end{equation}
where
\begin{equation}
A =\frac{2n+1-sw}{4ns}, \quad B=\frac{sw-1}{4n(n+1)s^{2n+1}}, \quad
C =-\frac{s(2n+1+sw)}{4(n+1)}, 
\end{equation}
with $n=1,2,\ldots$; $s$ is the half-thickness of the disc and $w$ is the jump of $h_{,zz}$ on $z=\pm s$. 
For $|z|>s$, we have $h_{,z}=\pm 1$ and $h_{,zz}=0$, so there is no matter in this region. We will 
consider the special case $w=0$, so $h_{,zz}(z=\pm s)=0$ and the density also vanishes there.
Figs.\ \ref{fig_rho_c0}(a)--(c) show isodensity curves in the three orthogonal coordinate planes 
of the mass density $\rho_{c(0)}/(M/b^3)$ for the 
polynomial with $n=1$, half-thickness $s/b=1$ and shift parameter $a/b=0.6$. We find that the mass density
is non-negative in the range $0<a/b \lesssim 0.62$, and when the half-thickness is reduced to $s/b=0.5$ the interval
becomes $0<a/b \lesssim 0.70$. The main qualitative difference
between our system and the one studied by Ciotti \& Marinacci \cite{cm08} for the Miyamoto-Nagai shifted disc lies in
the finite thickness of the former.

\begin{figure}
\centering
\includegraphics[scale=0.75]{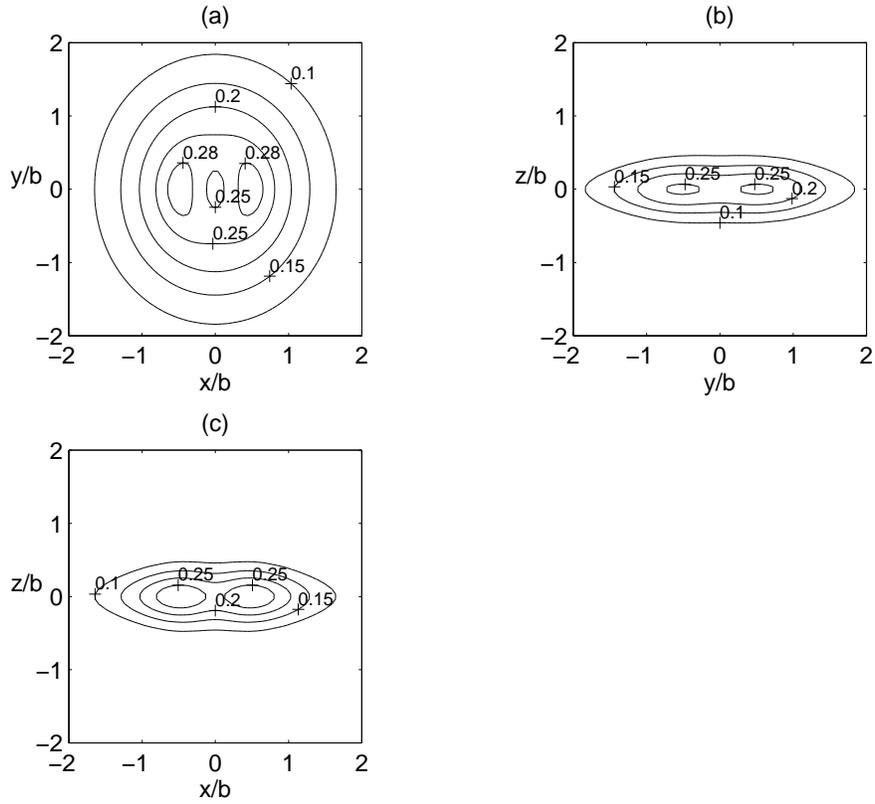}
\caption{Isodensity curves in the three orthogonal coordinate planes 
of the surface density $\rho_c^{(1,0)}/(\sigma_l/b)$ equation 
(\ref{eq_rho10}) for shift parameter $a/b=0.5$ and a Miyamoto-Nagai transformation 
with parameter $c/b=0.5$.} \label{fig_rho10}
\end{figure}
The ``inflated'' version of the ring-like non-axisymmetric flat systems (\ref{eq_sigma10}) and (\ref{eq_phi10}) also can 
be found by replacing $|z|$ by $h(z)$ in the potential (\ref{eq_phi10}) and calculating the resulting mass density 
with Poisson equation. One ends with
\begin{multline}
\rho_c^{(1,0)}=\frac{\sigma_lb^2h_{,zz}\sqrt{\zeta+\sqrt{\zeta^2+4a^2x^2}}}{6\sqrt{2}\left( \zeta^2+4a^2x^2 \right)^{5/2}}
\left\{ \left[ 3b+h(2+a^2/b^2) \right] \left( \zeta^2+4a^2x^2 \right) \right. \\ 
\left. \times \left( 2\zeta - \sqrt{\zeta^2+4a^2x^2} \right)  -3b(b+h)^2(1-a^2/b^2) 
\left( 3\zeta^2-2\zeta  \sqrt{\zeta^2+4a^2x^2} -4a^2x^2 \right) \right\} \\
+\frac{\sigma_lb^2\left(1-h_{,z}^2 \right)\sqrt{\zeta+\sqrt{\zeta^2+4a^2x^2}}}{6\sqrt{2}\left( \zeta^2+4a^2x^2 \right)^{7/2}}
\left\{ 3(b+h)\left[ 5b+2h+(h-2b)a^2/b^2 \right] \right. \\
\left. \times \left( \zeta^2+4a^2x^2 \right)
\left( 3\zeta^2-2\zeta  \sqrt{\zeta^2+4a^2x^2} -4a^2x^2 \right)  -15b(b+h)^3(1-a^2/b^2) \right. \\
\left. \times \left( 4\zeta^3-3\zeta^2\sqrt{\zeta^2+4a^2x^2}+4a^2x^2\sqrt{\zeta^2+4a^2x^2} 
 -16\zeta a^2x^2 \right) \right. \\
\left. -(2+a^2/b^2) \left( \zeta^2+4a^2x^2 \right)^2
\left( 2\zeta - \sqrt{\zeta^2+4a^2x^2} \right) \right\} \mbox{.} \label{eq_rho10}
\end{multline}
Some isodensity curves in the three orthogonal coordinate planes of the mass density $\rho_c^{(1,0)}/(\sigma_l/b)$ 
are shown in Figs.\ \ref{fig_rho10}(a)--(c), for the case of a Miyamoto-Nagai transformation $h(z)=\sqrt{z^2+c^2}$. 
The values of the parameters used were $a/b=0.5$ and $c/b=0.5$. As expected, one has a non-axisymmetric
toroidal mass distribution.   
\section{Using the imaginary part of complex-shifted discs} \label{sec_im}

\begin{figure}
\centering
\includegraphics[scale=0.7]{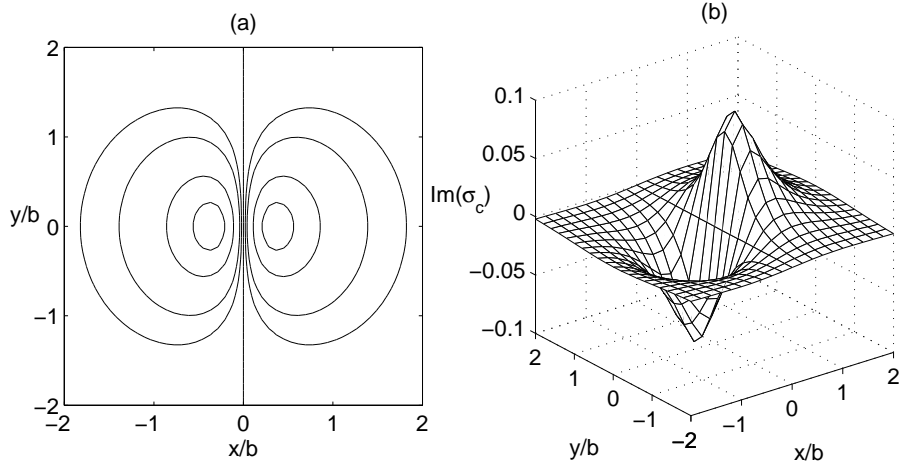}
\caption{(a) Isodensity curves and (b) surface plot of the
surface density $\Im(\sigma_c)/(M/b^2)$ equation (\ref{eq_im_sk}) with shift
parameter $a/b=0.5$.} \label{fig_im_s}
\end{figure}

In Section \ref{sec_complex}, we commented that the imaginary part 
of a complex-shifted density always changes sign \cite{cg07}, and therefore 
is not adequate for modelling gravitating systems. In fact, the imaginary part 
of the shifted Kuzmin-Toomre disc (\ref{eq_im_sk}) is negative for $x<0$. 
Figs.\ \ref{fig_im_s}(a) and (b) display level curves and the surface plot of the 
imaginary part of the dimensionless surface density with shift parameter $a/b=0.5$.

Now the combination of the imaginary part of a complex-shifted system with 
another potential-density pair may generate well-behaved and interesting 
new configurations. To illustrate our point, we take the superposition of 
the real and the imaginary part of the complex-shifted Kuzmin disc (\ref{eq_kuz_c})
\begin{equation} 
\sigma_{sup.} = \Re(\sigma_c)+\beta \Im(\sigma_c) \mbox{,}
\end{equation}
where $\beta$ is a dimensionless parameter which we assume positive, 
and an analogous expression holds for the potential. We have chosen this 
example for its simplicity, but in principle the superposition does not need to involve 
the real and imaginary parts of the same shifted system. Using equations 
(\ref{eq_re_pk})--(\ref{eq_im_sk}), we obtain 
\begin{gather} 
\Phi_{sup.} =-\frac{GM\left( \chi+\sqrt{\chi^2+4a^2x^2} +2\beta ax \right)}
{\sqrt{2}\sqrt{\chi^2+4a^2x^2}\sqrt{\chi+\sqrt{\chi^2+4a^2x^2}}} \mbox{,} \label{eq_sup_phi} \\
\sigma_{sup.} =\frac{Mb}{2 \sqrt{2} \pi \left( \xi^2+4a^2x^2 \right)^{3/2}
\sqrt{\xi+\sqrt{\xi^2+4a^2x^2}}}  \left[ \left( \xi+\sqrt{\xi^2+4a^2x^2} \right) \right. \notag \\
\left. \times \left( 2\xi-\sqrt{\xi^2+4a^2x^2} \right) +2\beta ax \left( 2\xi+\sqrt{\xi^2+4a^2x^2} \right) \right] \mbox{.} \label{eq_sup_s}
\end{gather}

\begin{figure}
\centering
\includegraphics[scale=0.75]{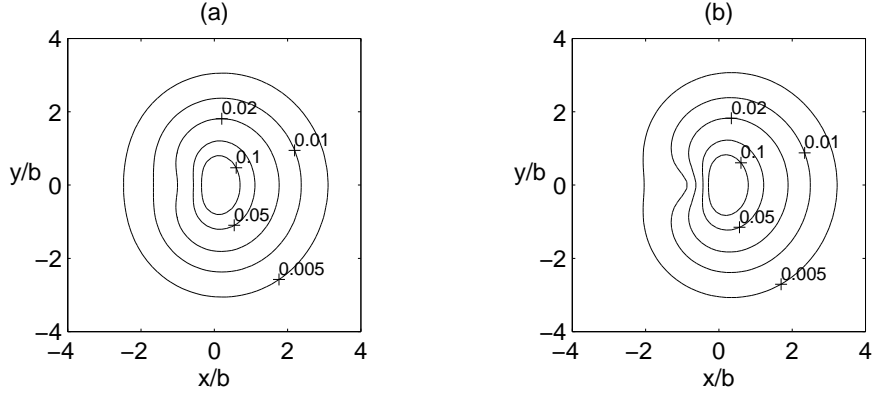}
\caption{Isodensity curves of the surface density $\sigma_{sup.}/(M/b^2)$ equation 
(\ref{eq_sup_s}) for shift parameter $a/b=0.5$ and (a) $\beta=0.5$, (b) $\beta=0.8$.} \label{fig_s_sup}
\end{figure}
Isodensity curves of (\ref{eq_sup_s}) are shown in Figs.\ \ref{fig_s_sup}(a) and (b) for
shift parameter $a/b=0.5$,  $\beta=0.5$ in Fig.\ \ref{fig_s_sup}(a) and $\beta=0.8$ in
Fig.\ \ref{fig_s_sup}(b). The mass distribution is no more symmetric with respect to 
the $x=0$ axis. For $x>0$, the imaginary part of the shifted density adds mass to the real 
part, and when $x<0$ there is a subtraction of mass. The maximum of density is displaced 
from the centre along the $x$-axis; in the examples shown in Fig.\ \ref{fig_s_sup}(a) the maximum is located at
$x/b \approx 0.09$ and in Fig.\ \ref{fig_s_sup}(b) at $x/b \approx 0.13$. For the values of $\beta$ shown
in Figs.\ \ref{fig_s_sup}(a) and (b), the surface density is non-negative everywhere, but one 
can expect that if $\beta$ becomes too large regions with negative densities should appear. In 
fact, we find the following intervals of $\beta$, depending on the shift parameter, where the 
superposed density is non-negative: for $a/b=0.25$, $0< \beta \lesssim 2.5$; for $a/b=0.5$,
$0< \beta \lesssim 1.0$; and for $a/b=0.75$, $0< \beta \lesssim 0.3$. 

\begin{figure}
\centering
\includegraphics[scale=0.75]{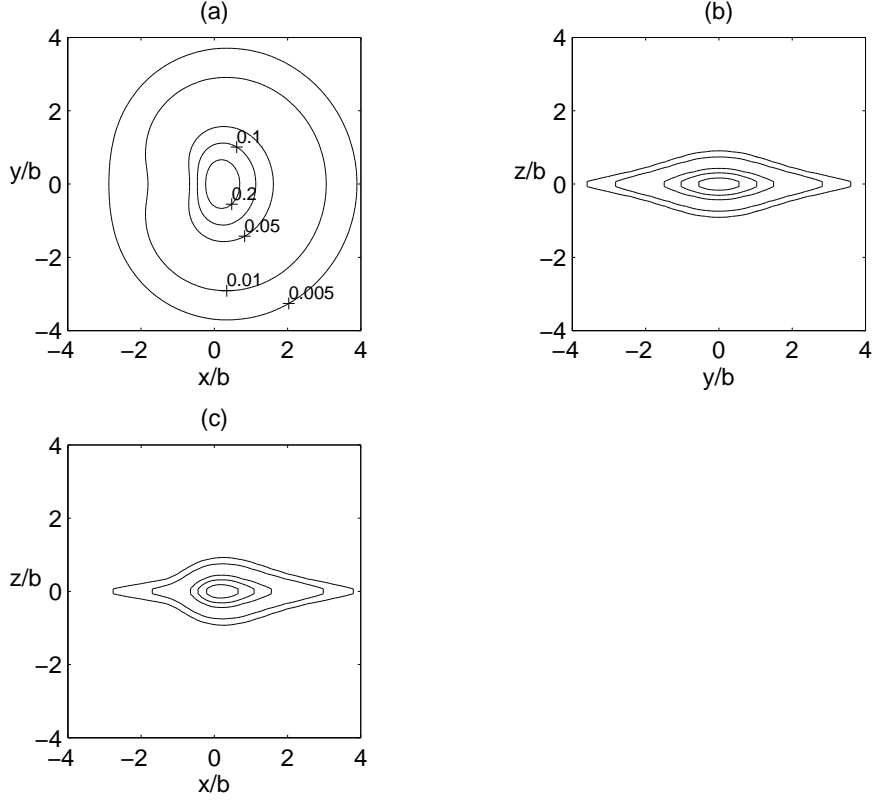}
\caption{Isodensity curves in the three orthogonal coordinate planes 
of the surface density $\rho_{sup.}/(M/b^3)$ equation 
(\ref{eq_rho_th}) for parameters $a/b=0.5$, $\beta=0.8$ and a Miyamoto-Nagai transformation 
with parameter $c/b=0.3$.} \label{fig_r_sup}
\end{figure}

It is also possible to ``inflate'' the flat potential-density pair 
(\ref{eq_sup_phi}) and (\ref{eq_sup_s}) by using the same procedure 
described in Section \ref{sec_tri_disc}. One obtains the following result: 
\begin{gather}
\Phi_{sup.}=-\frac{GM\left( \zeta+\sqrt{\zeta^2+4a^2x^2} +2\beta ax \right)}
{\sqrt{2}\sqrt{\zeta^2+4a^2x^2}\sqrt{\zeta+\sqrt{\zeta^2+4a^2x^2}}} \mbox{,} \\
\rho_{sup.}= \frac{M(b+h)h_{,zz}}{4\sqrt{2}\pi \left( \zeta^2+4a^2x^2 \right)^{3/2}\sqrt{\zeta+\sqrt{\zeta^2+4a^2x^2}}}
\left[ \left( \zeta+\sqrt{\zeta^2+4a^2x^2} \right) \right. \notag \\ 
\left. \times \left( 2\zeta-\sqrt{\zeta^2+4a^2x^2} \right)+2\beta ax \left( 2\zeta+\sqrt{\zeta^2+4a^2x^2} \right) \right] \notag \\
+ \frac{M\left(1-h_{,z}^2 \right)}{4\sqrt{2}\pi \left( \zeta^2+4a^2x^2 \right)^{5/2}\sqrt{\zeta+\sqrt{\zeta^2+4a^2x^2}}}
\left\{ -\left( \zeta^2+4a^2x^2 \right) \right. \notag \\
\left. \times \left[ \left( \zeta+\sqrt{\zeta^2+4a^2x^2} \right) \left( 2\zeta-\sqrt{\zeta^2+4a^2x^2} \right)
 + 2\beta ax \left( 2\zeta+\sqrt{\zeta^2+4a^2x^2} \right) \right] \right. \notag \\
\left. +3(b+h)^2 \left[  \left( \zeta+\sqrt{\zeta^2+4a^2x^2} \right) 
\left( 3\zeta^2-2\zeta \sqrt{\zeta^2+4a^2x^2} -4a^2x^2 \right) \right. \right. \notag \\
\left. \left. + 2\beta ax \left(  3\zeta^2+2\zeta \sqrt{\zeta^2+4a^2x^2} -4a^2x^2 \right) \right] \right\} \mbox{,} \label{eq_rho_th}
\end{gather} 
where, as before, $\zeta=x^2+y^2-a^2+[b+h(z)]^2$. As an example we take $h(z)=\sqrt{z^2+c^2}$. 
Figs.\ \ref{fig_r_sup}(a)--(c) show isodensity 
curves of the mass density $\rho_{sup.}/(M/b^3)$ on the three orthogonal coordinate planes
for  parameters $a/b=0.5$, $\beta=0.8$ and $c/b=0.3$. In Figs.\ \ref{fig_r_sup}(b) and (c), the values of the level 
curves are the same as indicated in Fig.\ \ref{fig_r_sup}(a). Again the lack of symmetry with respect to the 
$x=0$ axis is noted.
\section{Discussion} \label{sec_discuss}

The method of complexification, also known as complex-shift method, was used 
to generate a family of non-axisymmetric flat discs using the Kuzmin-Toomre family of discs 
as the parent system. The discs present  non-negative surface densities for restricted 
ranges of the shift parameter. Further we showed how these discs 
can be superposed to yield non-axisymmetric structures with vanishing surface density 
at the origin that can be considered as complex-shifted flat rings. We also ``inflated'' the  
aforementioned flat potential-density pairs by a general transformation and presented 
some examples by using a Miyamoto-Nagai transformation and a class of polynomials. 
These resulted in triaxial potential-density pairs. We also showed that the imaginary part 
of a complex-shifted system, which has non-physical characteristics, can be combined 
with other pairs to generate well-behaved systems that are not symmetrical with respect to
a coordinate axis. 

The analytical potential-density pairs presented in this work can all be expressed in 
terms of elementary functions, and yet show non-trivial features. We hope they may be useful 
as more realistic models in galactic dynamics. 

\section*{Acknowledgments}

We thank FAPESP for financial support, PSL also thanks CNPq. This research has 
made use of SAO/NASA's Astrophysics Data System abstract service, which is gratefully 
acknowledged.


\begin{thebibliography}{99}
\bibitem{bt08} Binney J., Tremaine S., 2008, Galactic Dynamics, 2nd edn. Princeton Univ. Press, Princeton, NJ
\bibitem{k56} Kuzmin G.G., 1956, Astron. Zh., 33, 27
\bibitem{ez92} Evans N.W., de Zeeuw P.T., 1992, MNRAS, 257, 152
\bibitem{blk92} Bi\v{c}\'{a}k J., Lynden-Bell D., Katz J., 1992, Phys. Rev. D, 47, 4334
\bibitem{blp93} Bi\v{c}\'{a}k J., Lynden-Bell D., Pichon C., 1993, MNRAS, 265, 126
\bibitem{lz99} Ledvinka T., Zofka M., Bi\v{c}\'{a}k J., 1999, in Piran T., ed.,   
Proc. 8th Marcel Grossman Meeting in General Relativity, World Scientific, Singapore, p. 339-341.
\bibitem{let99} Letelier P.S, 1999, Phys. Rev. D, 60, 104042
\bibitem{gl00} Gonz\'alez G.A., Letelier P.S., 2000, Phys. Rev. D, 62, 064025
\bibitem{vl03} Vogt D., Letelier P.S., 2003, Phys. Rev. D, 68, 084010
\bibitem{gl04} Gonz\'alez G.A., Letelier P.S., 2004, Phys. Rev. D, 69, 044013
\bibitem{vl05} Vogt D., Letelier P.S., 2005, Phys. Rev. D, 71, 084030
\bibitem{ap87} Appell P., 1887, Ann. Math. Lpz., 30, 155
\bibitem{ww50} Whittaker E.T., Watson G.N., 1950, A Course of Modern Analysis. Cambridge Univ. Press, Cambridge
\bibitem{lo87} Letelier P.S, Oliveira S.R., 1987, J. Math. Phys., 28, 165
\bibitem{gp89} Gleiser R.J., Pullin J.A., 1989, Class. Quantum Gravity, 6, 977
\bibitem{lo98} Letelier P.S, Oliveira S.R., 1998, Class. Quantum Gravity, 15, 421
\bibitem{dlo05} D'Afonseca L.A., Letelier P.S, Oliveira S.R., 2005, Class. Quantum Gravity, 22, 3803
\bibitem{cg07} Ciotti L., Giampieri G., 2007, MNRAS, 376, 1162
\bibitem{cm08} Ciotti L., Marinacci F., 2008, MNRAS, 387, 1117
\bibitem{t63} Toomre A., 1963, ApJ, 138, 385
\bibitem{nm76} Nagai R., Miyamoto M., 1976, PASJ, 28, 1
\bibitem{vl09} Vogt D., Letelier P.S., 2009, MNRAS, 396, 1487
\bibitem{ts76} Theys J.C., Spiegel E.A., 1976, ApJ, 208, 650
\bibitem{mn75} Miyamoto M., Nagai R., 1975, PASJ, 27, 533
















\end{thebibliography}
\end{document}